\documentclass[pdftex,twocolumn,epjc3]{svjour3}

\RequirePackage[T1]{fontenc}
\RequirePackage{graphicx}
\RequirePackage{mathptmx}
\RequirePackage{flushend}
\RequirePackage[numbers,sort&compress]{natbib}
\RequirePackage[colorlinks,citecolor=blue,urlcolor=blue,linkcolor=blue]{hyperref}

\usepackage[below]{placeins}
\usepackage{amsmath}
\usepackage{amssymb}
\usepackage{mathtools}
\usepackage{enumitem}
\usepackage{overpic}
\usepackage{tabularx}
\usepackage{siunitx}
\usepackage{stfloats}
\usepackage{microtype}
\usepackage{orcidlink}
\DeclareMathAlphabet{\mathcal}{OMS}{cmsy}{m}{n}

\newcommand{\SSD}{\textit{Solid\-State\-Detectors.jl}}
\usepackage{tikz}
\usetikzlibrary{positioning}
\usetikzlibrary{calc}
\definecolor{mylightred}{RGB}{211,79,73}
\definecolor{mydarkred}{RGB}{199,44,38}
\definecolor{mylightgreen}{RGB}{78,153,67}
\definecolor{mydarkgreen}{RGB}{43,129,33}
\definecolor{mylightpurple}{RGB}{150,107,178}
\definecolor{mydarkpurple}{RGB}{126,78,160}
\definecolor{mylightblue}{RGB}{49,101,205}
\definecolor{mydarkblue}{RGB}{20,92,205}
\tikzset{juliadot/.style args={#1,#2}{shape=circle,line width=0.03ex,minimum width=0.4ex,fill=#1,draw=#2}}
\newcommand\julialetter[1]{{\strut\fontfamily{cmss}\bfseries\selectfont{#1}}}
\DeclareRobustCommand\julia{%
\begin{tikzpicture}[baseline=0mm, every node/.style={inner sep=0mm, outer sep=0mm}]
\node[anchor=base]        (j) at (0,0) {\julialetter{\j}};
\node[anchor=base, right=0ex of j] (u) {\julialetter{u}};
\node[anchor=base, right=0ex of u] (l) {\julialetter{l}};
\node[anchor=base, right=0ex of l] (i) {\julialetter{\i}};
\node[anchor=base, right=0ex of i] (a) {\julialetter{a}};
\path let \p1 = (j) in node[juliadot={mylightblue,mydarkblue}] (bluedot) at (\x1+0.02ex,1.4ex) {};
\path let \p1 = (i) in node[juliadot={mylightred,mydarkred}] (reddot) at (\x1,1.4ex) {};
\path let \p1 = (reddot) in node[juliadot={mylightpurple,mydarkpurple}] (purpledot) at (\x1+0.5ex,\y1) {};
\path let \p1 = (reddot) in node[juliadot={mylightgreen,mydarkgreen}] (greendot) at (\x1+0.25ex,\y1+0.42ex) {};
\end{tikzpicture}%
}

\usepackage{lipsum}
\usepackage[switch]{lineno}
\nolinenumbers
\journalname{Eur. Phys. J. C}

\begin{document}

\title{Three-dimensional Compton imaging of undepleted volumes of germanium detectors}

\author{Iris~Abt\thanksref{MPP}\orcidlink{0009-0003-8821-5048}
\and Arthur~Butorev\thanksref{MPP}\orcidlink{0009-0003-9655-4904}
\and Felix~Hagemann\thanksref{MPP,SSL,e}\orcidlink{0000-0001-5021-3328}
\and David~Hervas Aguilar\thanksref{MPP,UNC,TUNL,TUM}\orcidlink{0000-0002-9686-0659}
\and Johanna~L\"uhrs\thanksref{MPP,MPIK}\orcidlink{0009-0004-8414-6206}
\and Julia~Penner\thanksref{MPP,TUM} \orcidlink{0009-0001-2980-9052}
\and Oliver~Schulz\thanksref{MPP}\orcidlink{0000-0002-4200-5905}
}
\thankstext{e}{e-mail: hagemann@mpp.mpg.de (corresponding author)}
\institute{Max-Planck-Institut f\"ur Physik, Boltzmannstra\ss{}e 8, 85748 Garching bei M\"unchen, Germany\label{MPP} 
\and
University of North Carolina, Department of Physics and Astronomy, 120 E. Cameron Ave., Phillips Hall CB3255, Chapel Hill, 27599 NC, USA\label{UNC} 
\and 
Triangle Universities Nuclear Laboratory, 116 Science Drive, Duke University, Durham, 27708 NC, USA\label{TUNL}
\and 
\emph{Present Address:} Space Sciences Laboratory, University of California, Berkeley, 7 Gauss Way, Berkeley CA 94720, USA\label{SSL}
\and 
\emph{Present Address:} Department of Physics, TUM School of Natural Sciences, Technical University of Munich, James-Franck-Stra\ss{}e 1, 85748 Garching bei M\"unchen, Germany \label{TUM}
\and
\emph{Present Address:} Max-Planck-Institut für Kernphysik, Saupfercheckweg 1, 69117 Heidelberg, Germany \label{MPIK}
}

\date{Received: date / Accepted: date}

\maketitle

\begin{abstract}

The shape of the undepleted volume of a p-type Broad Energy Germanium detector has been imaged at various bias voltages by measuring spatially-resolved three-dimensional Compton-scattering efficiency.
The bias voltage was raised stepwise from $-50\,\si{\volt}$ to the full-depletion voltage.
The geometric acceptance was determined at full depletion. 
Below full depletion, the relative acceptance observed for $2\times2\times2\,\si{\milli\meter}^3$ voxels was used to create the image of the undepleted volume for each bias voltage. 
The images were fitted with predictions from the open-source software package \SSD{} to extract the radial impurity density profile of the detector. 
The strong radial decrease in the impurity density obtained in the outermost centimeter was validated using independent capacitance measurements. 
This novel method to determine the impurity density via three-dimensional Compton-scattering efficiency maps provides a powerful tool for advanced detector bulk characterization.

\end{abstract}

\section{Introduction}

High-purity germanium detectors are well known for their efficiency in detecting gamma rays with excellent energy resolution. 
This is sufficient for many applications in industry and research.
However, for rare-event searches like the search for neutrinoless double-beta decay~\mbox{\cite{GERDA_PSA, Majorana_PSA, LEGEND}} or for dark matter~\mbox{\cite{SuperCDMS, CDEX, EDELWEISS}}, more information is needed. 
Additional information is deduced from the shape of the charge pulse produced by a germanium detector in response to the interaction of a particle with the crystal. 
This so-called pulse shape analysis is used to separate signal events from background and relies strongly on verification from comparisons between measured and simulated pulses.
Thus, the basic physics inside a germanium detector as well as the parameters of the simulated detector have to be extremely well understood.
 
In order to operate a germanium detector, a bias voltage, $V_B$, is applied to the contacts of the detector to generate an electric field inside the detector. 
This field has two functions:
\begin{enumerate}
\item It sweeps out the free charge carriers produced inside the crystal by the impurities; this process is called depletion, during which the electrically active impurities are ionized. These ionized impurities also affect the electric field.
\item It separates the electrons and holes produced by particle interactions; this process generates the desired signal.
\end{enumerate}
The electric field is confined to the depleted volume which, thus, defines the active volume of the detector.
A detector is fully operational when it is depleted throughout its volume. 
This happens for a voltage $V_B$ above the so-called full-deple\-tion voltage, $V_D$.
For lower $V_B$, efficient charge collection is expected only in the depleted volume.
 
The simulation of the pulses of a detector requires a realistic calculation of the electric field. 
This, in turn, requires the impurity density profile of the detector as an important input.
However, the impurity density profile is a priori not well known. 
The manufacturer typically provides impurity values at the top and the bottom of the crystal, usually obtained via Hall measurements with associated uncertainties of approximately 20\,\%~\cite{Mertens2019}.
In general, no radial dependence is assumed.
The values of $V_D$ as calculated based on the provided impurity density values and a linear interpolation between them often do not match the observed values of $V_D$ with deviations of up to 40\,\%.
In addition, some radial dependencies of the impurity density have been observed lately\,\mbox{\cite{Abt2023,Mei2016,Mei2017}}.

As the evolution of the undepleted volume with increasing $V_B$ depends on the impurity density profile, the observation of the way the undepleted volume shrinks can be used to obtain information about the impurity density profile.
Historically, imaging undepleted volumes was restricted to integrated projections.
For example, the depletion behavior of segmented AGATA detector capsules was investigated using reduced bias voltage scans~\cite{Colosimo2015}, where the detectors were irradiated from the top and the side with a collimated beam from a $^{137}$Cs source.
Rather than resolving individual interaction positions along the beam axis, the total count rates of single-segment photopeak events at different source positions were used to form two-dimensional projections.

In contrast, the wide-angle Compton Scanner at the Max-Planck-Institute for Physics~\cite{Abt2022} reconstructs events back to specific points along the beam axis.
The special feature of the Compton Scanner is that it can probe all $z$ positions in a detector simultaneously. 
A first proof of principle demonstrated the feasibility of using the Compton Scanner to compose three-dimensional images of the undepleted volumes of a multi-kilogram inverted-coaxial point-contact germanium detector~\cite{Hervas2023}.
While limited statistics did not allow for the extraction of a radial impurity density profile, these images were sufficient to infer a longitudinal $z$-gradient.

In this paper, a novel method to extract the full impurity density profile of a high-purity germanium detector based on the three-dimensional imaging capability of the Compton Scanner is presented.
It is shown that a radial dependence of the impurity density profile is needed to explain the images. 
This result is compared to results obtained from capacitance measurements. 
In the conclusion, we present suggestions on how germanium detectors should be characterized before being deployed in large experiments.

\section{The p-type segBEGe detector}
\label{sec:det}

\begin{figure*}[!p]
    \centering
    \begin{overpic}[width =1.05\textwidth]{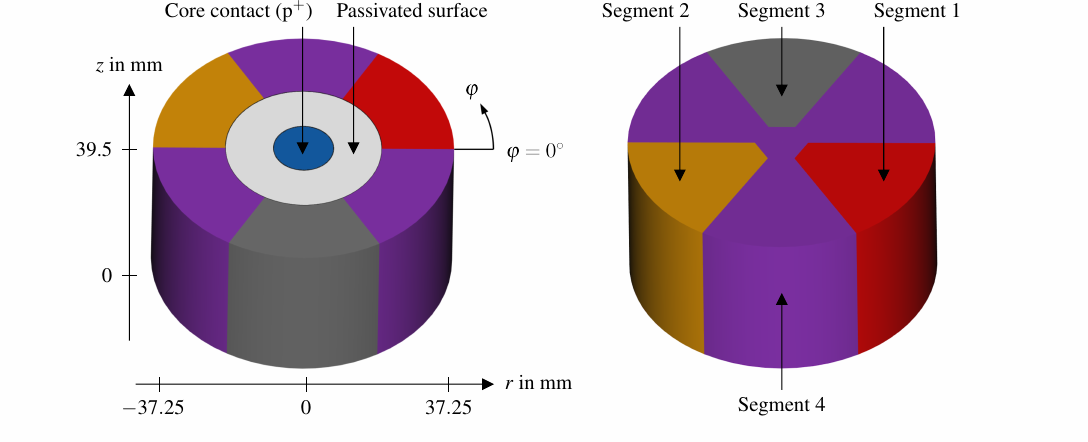} 
    \bfseries\large
    \put(28.1,0.5){\makebox(0,0){(a)}}
    \put(71.9,0.5){\makebox(0,0){(b)}}
    \end{overpic}
    \caption{
    Schematic of the p-type segBEGe detector: (a) top view and (b) bottom view. 
    In (a), the cylindrical coordinate system to describe positions in the segBEGe detector is shown. 
    The segments had lithium-drifted contacts which were grounded for the data taken for this paper.
    }
    \label{fig:segBEGe}
\end{figure*}

The detector used for this study was a p-type segmented Broad Energy Germanium (segBEGe) detector, manufactured by Mirion Technologies.
It was a cylindrical detector with a diameter of \mbox{$74.5\,\si{\milli\meter}$} and a height of \mbox{$39.5\,\si{\milli\meter}$}, see Fig.~\ref{fig:segBEGe}. 
The p$^+$~point contact on top of the detector, the so-called core contact, was a boron implant. 
It had a diameter of \mbox{$15\,\si{\milli\meter}$} and was surrounded by a passivated ring with an outer diameter of \mbox{$39\,\si{\milli\meter}$}. 
The bottom, side and outer part of the top surface of the detector were covered with a four-fold segmented lithium-drifted n-type contact. 
The thickness of the lithium contact at the side of the detector was measured in radial scans to be \mbox{$0.75\pm0.06\,\si{\milli\meter}$}~\mbox{\cite[p.\,123]{Hagemann2024}}. 
The smaller segments~1, 2~and~3 each extended over \mbox{$60\si{\degree}$}, while segment~4 covered the remaining surface on the top and side of the detector and had a closed bottom end-plate. 
However, the segmentation is not important for the analyses presented in this paper because the contacts were grounded and the detector was effectively a BEGe detector.

Figure~\ref{fig:segBEGe}a also depicts the cylindrical coordinate system, $r$, $\varphi$ and $z$, used to describe positions in the detector. 
The origin of the coordinate system is located in the center of the bottom surface of the detector. 
The segment boundary between segment~1 and~4 which was closer to segment~3 defines \mbox{$\varphi = 0\si{\degree}$}.
The bottom surface of the detector defines \mbox{$z = 0\,\si{\milli\meter}$} with the $z$-values increasing towards the top of the detector at \mbox{$z = 39.5\,\si{\milli\meter}$}.

\begin{figure*}[!p]
    \centering
    \includegraphics[width=1.05\textwidth]{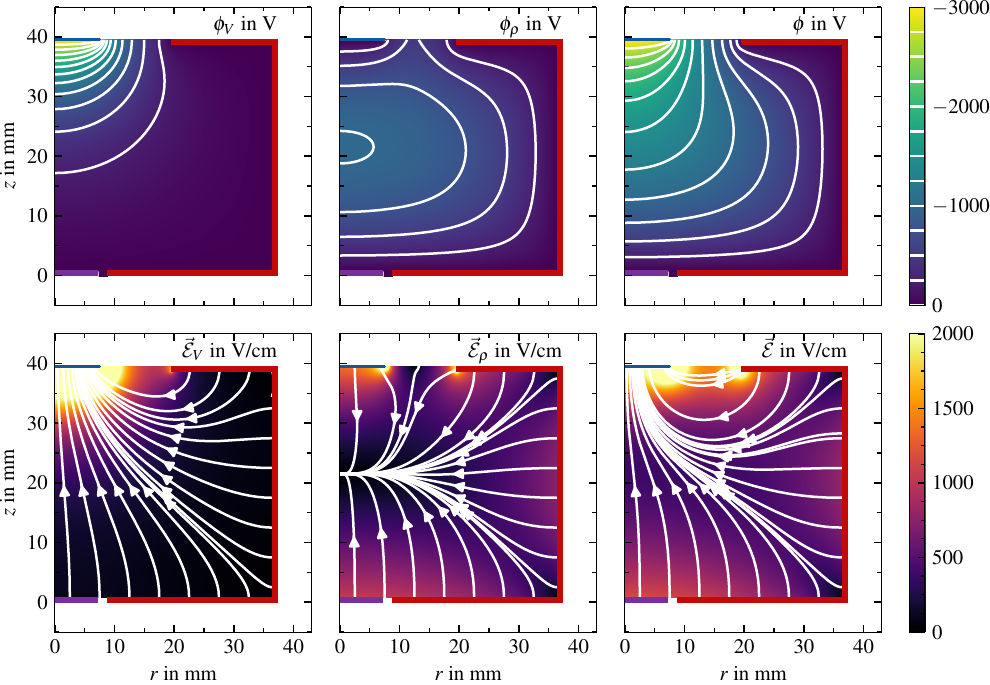} 
    \caption{Top row: electric potential in the $r$-$z$-plane at \mbox{$\varphi = 30^\circ$} resulting from (left) applying just the potentials to the contacts, $\phi_V$, (center) just the charge density from ionized impurities, $\phi_\rho$ and (right) total electric potential, $\phi$, as calculated with \SSD~\cite{Abt2021}, assuming a linear impurity density profile with the detector manufacturer values scaled to 89\%, i.e.\;\mbox{\smash{$\rho_\text{top} = -5.79\cdot10^9\,\si{\per\centi\meter\cubed}$}} and 
    \mbox{\smash{$\rho_\text{bot} = -5.16\cdot10^9\,\si{\per\centi\meter\cubed}$}}. The white lines depict equipotential lines at integer multiples of $-250$\,V.
    Bottom row: corresponding electrical fields. The white lines represent field lines.}
    \label{fig:epot-components}
\end{figure*}

The detector manufacturer~\cite{Mirion} reported the net impurity density of the p-type segBEGe detector to be 12.1\,\% 
higher at the top, i.e.\;\mbox{\smash{$\rho_\text{top} = -6.5\cdot10^{9}\,\si{\per\centi\meter\cubed}$}}, compared to at the bottom of the detector, i.e.\;$\rho_\text{bot} = -5.8\cdot10^9\,\si{\per\centi\meter\cubed}$.
The calculation using these values for the impurity densities, assuming a linear $z$-dependence%
\footnote{For the moderate height of $39.5\,\si{\milli\meter}$ and the relatively low difference of 12.1\,\%, the assumption of linearity is reasonable. For larger-height detectors or larger differences, better models might be needed~\cite{Radford2025}.}
and no $r$-dependence, predicted a depletion voltage of \mbox{$V_D = -1437\,\si{\volt}$}.
To reproduce the lower measured value of \mbox{$V_D = -1275\,\si{\volt}$}~\cite{Hagemann2024}, the impurity density values had to be scaled down to 89\,\%, i.e.\;to \mbox{\smash{$\rho_\text{top} = -5.79\cdot10^9\,\si{\per\centi\meter\cubed}$}} and \mbox{\smash{$\rho_\text{bot} = -5.16\cdot10^9\,\si{\per\centi\meter\cubed}$}}.
The standard operating voltage was \mbox{$-3000\,\si{\volt}$} to assure efficient charge collection.
The calculated electric potential and electric field inside the detector is presented in Fig.~\Ref{fig:epot-components} and explained in section~\ref{sec:sim}.

\section{Experimental setup}
\label{sec:setup}

\begin{figure*}[!tb]
    \centering
    \begin{overpic}[width = \textwidth]{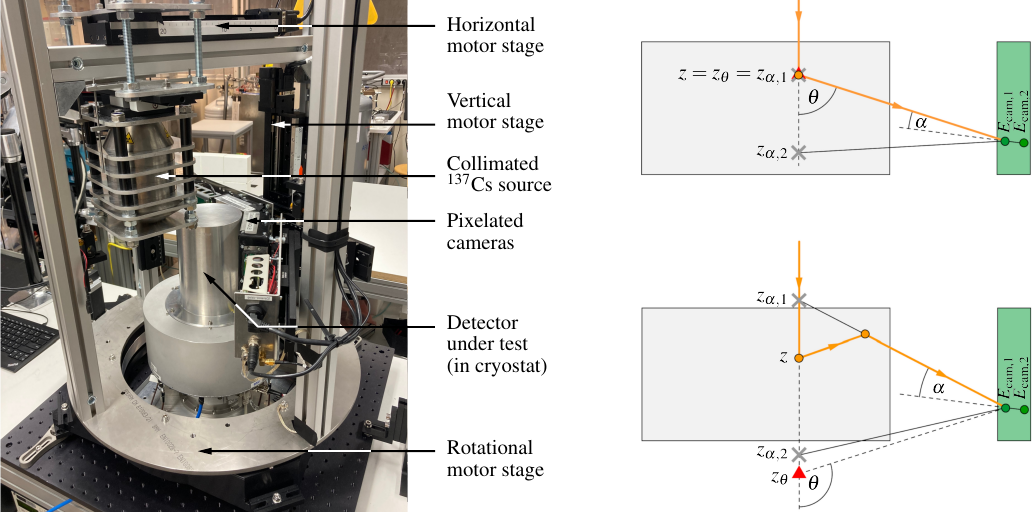}
    \bfseries\large
    \put(19.7,-1.8){\makebox(0,0){(a)}}
    \put(77.5,30.8){\makebox(0,0){(b)}}
    \put(77.5,-1.8){\makebox(0,0){(c)}}
    \end{overpic}
    \smallskip
    \label{fig:Cscanner} 
    \caption{
    (a) Photo of the Compton Scanner setup. Compared to the original setup~\cite{Abt2022}, a second camera was added to increase the acceptance of scattered gammas.
    Also shown are schematics of the event reconstruction for (b)~a single-site event and (c)~a multi-site event in the germanium detector~\cite{Abt2022}.
    Dots denote true interaction positions in the detector and camera.
    Triangles represent the positions $z_\theta$ reconstructed using the energy ratios between detector and camera.
    Crosses indicate spatial candidates derived from intersections of the camera-based Compton cone with the beam axis. In (b), the dot, the triangle and one of the crosses~overlap.
    }
\end{figure*}

The measurements for this paper were taken with the fully automated Compton Scanner~\cite{Abt2022} shown in Fig.~\ref{fig:Cscanner}a. 
The setup was developed and operated at the Max-Planck-Institute for Physics to characterize the bulk of germanium detectors.
It relies on reconstructing interaction positions from single-Compton-scattering events measured in coincidence with surrounding pixelated cameras, which allows pulse shapes from individual events to be mapped to their spatial origin.

The p-type segBEGe detector was cooled to \mbox{$78\,\si{\kelvin}$} in the temperature-controlled cryostat K2~\mbox{\cite[p.\,69]{Hagemann2024}} placed at the center of the Compton Scanner.
The detector was biased by grounding all segment contacts and applying the negative potential $V_B$ to the core contact.
The charges induced on the core contact were amplified using a Canberra PSC823C charge-sensitive preamplifier
and recorded using the STRUCK SIS3316-250-14 analog-to-digital converter (ADC) with a sampling time of $4\,\si{\nano\second}$ and a $14$\,bit resolution. 

The detector was irradiated with a collimated beam of \mbox{$661.66\,\si{\kilo\electronvolt}$} gammas from a \mbox{$740\,\si{\mega\becquerel}$} $^{137}$Cs source with an aperture angle of 0.26$^\circ$.
The beam spot diameter at the top (bottom) of the detector was \mbox{$1.23\,\si{\milli\meter}$} (\mbox{$1.61\,\si{\milli\meter}$}) for all radii, resulting in an average size of the beam spot of \mbox{$1.40\,\si{\milli\meter}$}. 
A system of motors was used to place the source at any position above the detector with a precision of better than \mbox{$100\,\si{\micro\meter}$}.

The gammas from Compton-scattering events were detected by a system of two pixelated cameras produced by H3D, Inc.\ placed at the side of the detector.
Each camera consists of a \mbox{$2\times2$} array of \mbox{$22\times22\times10\,\si{\milli\meter\cubed}$} CdZnTe crystals. 
The anode of each crystal is segmented into $11\times11$ pixels with a pixel pitch of $1.9\,\si{\milli\meter}$, yielding a position resolution of $0.5\,\si{\milli\meter}$ FWHM as specified by the manufacturer.

Events were selected if the germanium detector and the pixelated cameras detected energies, in coincidence, adding up to the characteristic energy of \mbox{$661.66\,\si{\kilo\electronvolt}$} within a window of \mbox{$\pm\,50\,\si{\kilo\electronvolt}$}.
The $x$ and $y$ positions of the Compton interaction were defined by the position of the $^{137}$Cs source.
The $z$ position along the beam axis was reconstructed using the topology of Compton scattering~\cite{Abt2022}.
Assuming a single Compton interaction in the germanium detector, a candidate $z_\theta$ is reconstructed based on the Compton scattering angle 
\begin{linenomath}
\begin{equation} \label{eq:compton}
\theta = \arccos\left(1 - \dfrac{m_ec^2 E_\text{det}}{\left(E_\text{det} + E_\text{cam}\right) E_\text{cam}}\right)~,
\end{equation}
\end{linenomath}
where $E_\text{det}$ is the total energy deposited in the detector and $E_\text{cam}$ is the total energy recorded in the camera, see Fig.~\ref{fig:Cscanner}b.

The assumption of a single Compton interaction in the germanium detector can be explicitly verified if the camera registers two hits, i.e.\;when the scattered gamma scatters once more in the camera before being photo-absorbed.
For these events, an additional Compton angle 
\begin{linenomath}
\begin{equation}
\alpha = \arccos\left(1 - \dfrac{m_ec^2 E_\text{cam,1}}{\left(E_\text{cam,1} + E_\text{cam,2}\right) E_\text{cam,2}}\right)
\end{equation}
\end{linenomath}
can be determined exclusively from the first and second camera hit, $E_\text{cam,1}$ and $E_\text{cam,2}$.
The positions of the two hits and this Compton angle $\alpha$ define a Compton cone which intersects the beam axis in up to two points, $z_\alpha$.
The event reconstruction is successfully validated only if $z_\theta$ agrees with one of these candidates within \mbox{$\vert z_\theta - z_\alpha\vert \leq 2\,\si{\milli\meter}$}.

Since the hit sequence in the camera is initially unknown, the lower-energy hit is first assigned to $E_\text{cam,1}$.
If the validation fails, the hit sequence is inverted.
If both attempts fail, the event is discarded as multi-site in the germanium detector, see Fig.~\ref{fig:Cscanner}c.
This validation method is highly effective at rejecting multi-site events, while reconstructing single-site events with an intrinsic $z$-resolution of \mbox{$\pm1\,\si{\milli\meter}$}~FWHM~\cite{Abt2022}.
On rare occasions, an incorrect hit sequence is falsely validated, resulting in large reconstruction errors and giving rise to Cauchy tails in the reconstructed $z$-distribution~\cite{Hagemann2024}.

For the analysis presented in this paper, only events that passed this validation were considered.
Furthermore,
the reconstructed $z_\theta$ is determined without relying on the energy measured by the detector, by setting \mbox{$E_\text{det} = 661.66\,\si{\kilo\electronvolt} - E_\text{cam}$} in Eq.\;\eqref{eq:compton}.
This is important for measurements at low $V_B$, where the energy resolution of the detector degrades.

\section{Simulation framework}
\label{sec:sim}

All potential and field calculations for this paper were performed with the open-source \julia{} simulation software package \SSD~\cite{Abt2021}. 

\SSD{} performs 2D and 3D calculations of the electric potential in the germanium detector using a successive over-relaxation algorithm on an adaptive grid with red-black division~\cite[p.\,57--64]{Hauertmann2021}, which allows efficient and parallel execution~\cite{Ortega1985}.
By grounding the segments, the p-type segBEGe detector becomes effectively $\varphi$-symmetric.
Full 3D calculations confirmed that the small gaps between the segments have a negligible impact on the electric field in the detector bulk.
Thus, all electric potential calculations for this paper were performed in 2D, which significantly accelerates the computation while preserving numerical accuracy.

The electric potential, $\phi(\vec{r})$, inside the detector is calculated using Gauss's law,~i.e.
\begin{linenomath}
\begin{align}
    \nabla \cdot (\epsilon_r(\vec{r}) \nabla \phi(\vec{r})) &= -\frac{\rho(\vec{r})}{\epsilon_0}, &\text{with} \quad\phi(\vec{r})\vert_{S_i} &= \phi_i~,
\end{align}
\end{linenomath}
where $\rho(\vec{r})$ is the density of space charge originating from ionized impurities,\footnote{No charge accumulation on surfaces was considered as it does not influence the bulk volumes discussed in this paper.} {\sisetup{per-mode = symbol}\mbox{$\epsilon_0 = 8.854\cdot10^{-12}\,\si{\farad\per\meter}$}} is the dielectric constant, $\epsilon_r(\vec{r})$ is the relative permittivity of the material at position $\vec{r}$, e.g.~$16$ for germanium~\cite[p.\,218]{Leo1994}. 
The boundary conditions denote that a potential $\phi_i$ is applied to the contact $S_i$, where $i$ is the index of the contact. 
For the data presented in this paper, \mbox{$\phi_i=0\,\si\volt$} for all segments.  
The overall potential $\phi(\vec{r})$ can be expressed as a sum of two components,
\begin{linenomath}
\begin{equation}
\phi(\vec{r}) = \phi_V(\vec{r}) + \phi_\rho(\vec{r})~,
\end{equation}
\end{linenomath}
where $\phi_V(\vec{r})$ and $\phi_\rho(\vec{r})$ denote the contribution of the applied potentials and the ionized impurities, respectively:
\begin{linenomath}
\begin{align}
    \nabla \cdot (\epsilon_r(\vec{r}) \nabla \phi_V(\vec{r})) &= 0, &\text{with} \quad\phi_V(\vec{r})\vert_{S_i} &= \phi_i~, \\
    \nabla \cdot (\epsilon_r(\vec{r}) \nabla \phi_\rho(\vec{r})) &= -\frac{\rho(\vec{r})}{\epsilon_0}, &\text{with} \quad\phi_\rho(\vec{r})\vert_{S_i} &= 0~.
\end{align}
\end{linenomath}
The electric potential $\phi(\vec{r})$ of the detector, as well as the two components $\phi_V(\vec{r})$ and $\phi_\rho(\vec{r})$ are shown in Fig.~\ref{fig:epot-components}. 
The resulting electrical fields are also shown.
The $\phi_\rho(\vec{r})$ component is dominant at the bottom  and side of the detector. 
At the top, the situation also changes significantly by $\phi_\rho(\vec{r})$, which counteracts $\phi_V(\vec{r})$ and, thus, essentially weakens the electric field close to the point contact.
This demonstrates how important the knowledge of the impurity density in these volumes is to understand the pulses coming from the detector.

During the calculation of the electric potential, \SSD{} also determines the undepleted volume of the detector~\mbox{\cite[p.\,55]{Hagemann2024}}.
For a fully depleted detector, no local maxima or minima are observed in the bulk of the detector. 
If the potential of any grid point is smaller (bigger) than the minimum (maximum) of the potential of all neighboring grid points, the associated volume is marked as undepleted and the charge density induced by ionized impurities is scaled down until this is no longer the case. 
This results in an individual scaling factor $f$ for all grid points in the undepleted volume. 
This is done in each iteration of the calculation of the potential until it converges towards a steady state, in which the potential becomes constant in the undepleted volume.

\begin{figure}[!tb]
    \centering
    \smallskip
    \includegraphics{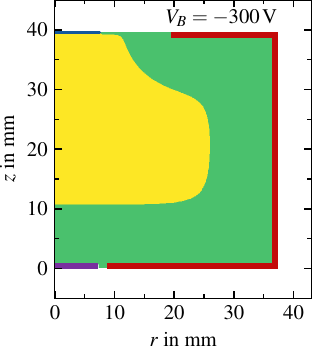}\qquad\null 
    \caption{Simulated undepleted volume (yellow) of the p-type segBEGe detector at $V_B=-300$\,V, using the default manufacturer values for the impurity density scaled to 89\,\% and a thickness of the lithium-drifted contacts of 0.75\,mm.}
    \label{fig:simulated-depletion}
\end{figure}

As an example, Fig.~\ref{fig:simulated-depletion} shows the extent of the undepleted volume of the detector as determined for \mbox{$V_B=-300\,\si{\volt}$}, using the values of the impurity density as provided by the manufacturer scaled to 89\,\% and assuming a thickness of \mbox{$0.75\,\si{\milli\meter}$} for the lithium-drifted contact.
The lower edge of the undepleted volume in $z$ is mostly determined by the $z$ profile of the impurity density profile. 
It was investigated that it is not very sensitive to deviations from the linear dependence assumed. 
Such deviations also do not change the $r$-extension significantly.
The depleted volume evolves from the p-n junction at the segment contacts towards the bulk of the detector. 
The undepleted volume shrinks as $V_B$ increases and, for this detector, is always connected to the core contact.

\section{Compton images} \label{sec:compton_images}

\begin{figure}[!t]
    \centering
    \vspace{-0.5\baselineskip}
    \includegraphics{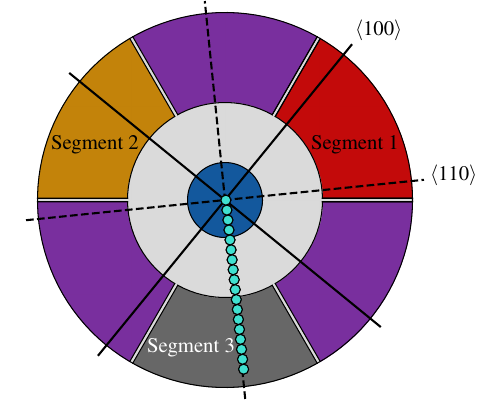} 
    \caption{Top view of the scan points used to image the undepleted volume of the p-type segBEGe detector. The solid and dashed black lines indicate the locations of the $\langle100\rangle$ and $\langle110\rangle$ axes, respectively.}
    \label{fig:scanplan}
\end{figure}

Figure~\ref{fig:scanplan} shows the positions of the source, at which data were taken with the Compton Scanner along the $\langle110\rangle$ axis in segment 3, starting at \mbox{$r=0\,\si{\milli\meter}$} and ending at \mbox{$r=34\,\si{\milli\meter}$}.

\begin{figure}[!t]
    \bigskip
    \centering
    \null\quad\begin{overpic}{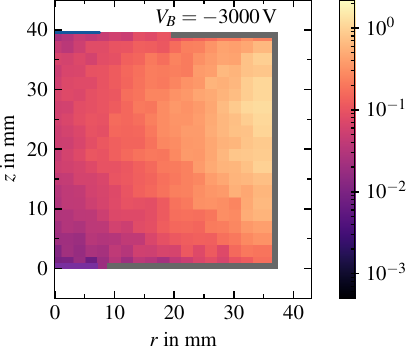} 
    \normalsize
    \put(90,88.5){\makebox(0,0)[lt]{cts/min}}
    \end{overpic}
    \caption{
    Measured rate of validated two-hit events for the detector operated at \mbox{$-3000\,\text{V}$} with logarithmic colorbar in counts/minute.
    The voxel sizes are $2\,\si{\milli\meter}$ in $r$ and $z$.
    }
    \label{fig:rate_depleted}
\end{figure}

The event rates in counts/minute for voxels of \mbox{$2\,\si{\milli\meter}$} in $z$ and $r$ for the standard operational $V_B$ of \mbox{$-3000\,\si{\volt}$} are shown in Fig.~\ref{fig:rate_depleted}. 
The voxel size of \mbox{$2\times2\times2\,\si{\milli\meter}^3$} reflects the intrinsic spatial resolution of the Compton Scanner of \mbox{$\pm\,1\,\si{\milli\meter}$} FWHM in $z$~\cite{Abt2022} and the full diameter of the beam spot in the $x$-$y$-plane.
The overall rates are low due to the requirement of two hits in the cameras.
The rates are lower at the bottom than at the top of the detector because of the absorption of incoming gammas.
The exponential decrease of the rate from the side to the center of the detector is due to the absorption of outgoing gammas and the acceptance of the cameras.
The measurement times were increased exponentially from 2 to 23\,hours towards the center of the detector to obtain  2400 validated two-hit events at each source position.
The total measurement time was about 7 days per~$V_B$.

Data with $\vert V_B\vert < \vert V_D\vert$ were taken at
$-50$, $-100$, $-150$, $-300$, $-600$ and \mbox{$-900\,\si{\volt}$}. 
As the energy resolution of the detector deteriorates at voltages significantly below $V_D$, the energy measured in the detector was only used to select events with fully-contained \mbox{$661.66\,\si{\kilo\electronvolt}$} gammas but not in the determination of the $z$ position of the Compton scattering.

\begin{figure*}[!p]
  \centering

  \begin{minipage}{\textwidth}
    \centering
    \begin{overpic}{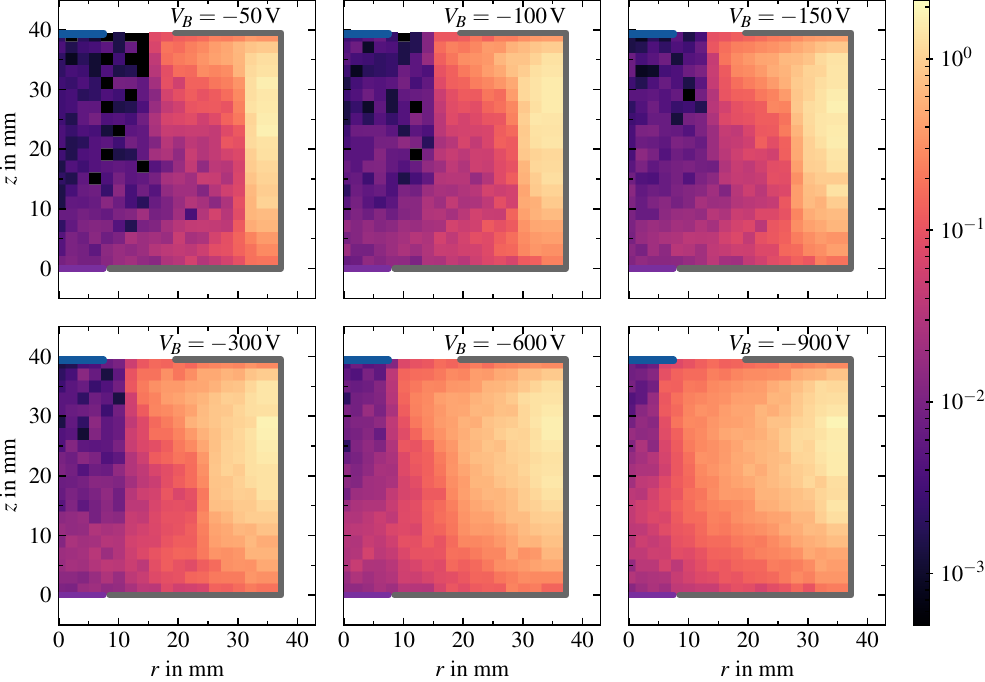} 
    \normalsize 
    \put(95.7,69.2){\makebox(0,0)[lt]{cts/min}}
    \end{overpic}
    \vspace{-0.25\baselineskip}
    \caption{Measured rates of validated two-hit events for $V_B$ between \mbox{$-50\,\si{\volt}$} and \mbox{$-900\,\si{\volt}$} with logarithmic colorbar. The voxel sizes are $2\,\si{\milli\meter}$ in $r$ and $z$.}
    \label{fig:rates_undepleted}
  \end{minipage}

  \vspace{\baselineskip}

  \begin{minipage}{\linewidth}
    \centering
    \includegraphics{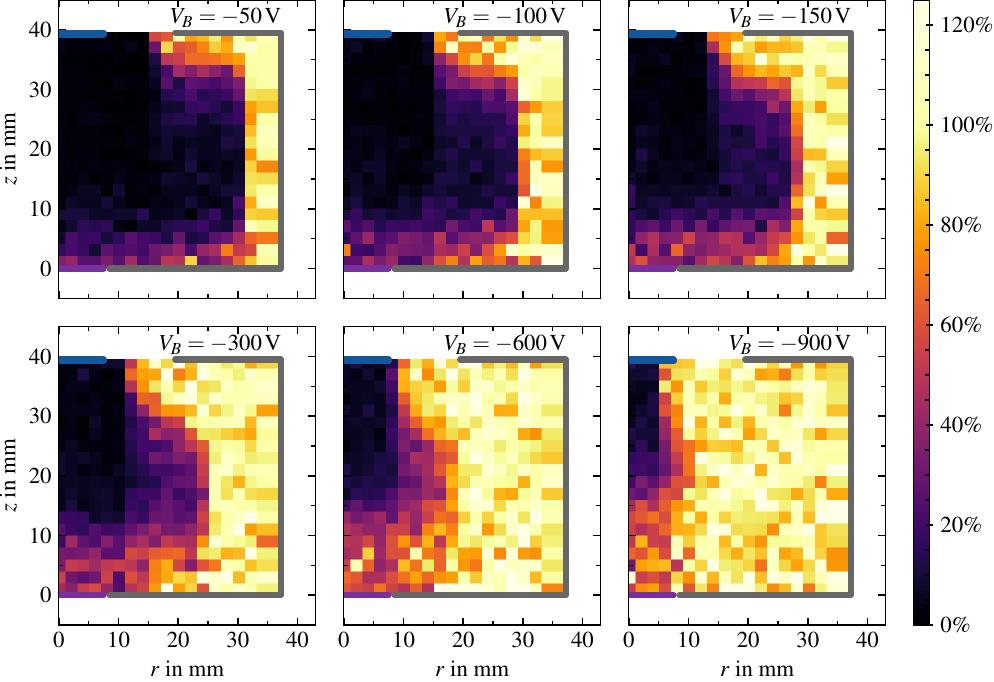} 
    \vspace{-0.25\baselineskip}
    \caption{Measured relative efficiencies, determined as the ratio of the rates from Fig.~\ref{fig:rates_undepleted} and the full-depletion rates at \mbox{$-3000\,\si{\volt}$}, see Fig.~\ref{fig:rate_depleted}. 
    The linear colorbar exceeds 100\,\% to show the statistical fluctuations in the region with $\sim100\,\%$ efficiency, which have a standard deviation $\sigma$ of~10\,\%.}
    \label{fig:bubbles_undepleted}
  \end{minipage}

\end{figure*}

The raw rates for the selected voltages are shown in Fig.~\ref{fig:rates_undepleted}.
Some regions exhibit significantly lower rates compared to the full-depletion rates at $-3000\,\si{\volt}$ shown in Fig.~\ref{fig:rate_depleted}.
Dividing the raw rates in Fig.~\ref{fig:rates_undepleted} by the full-depletion reference yields the relative efficiencies shown in Fig.~\ref{fig:bubbles_undepleted}.
Volumes with low relative efficiencies indicate poor charge collection and are, thus, identified as undepleted.
The volumes with high relative efficiency are identified as depleted.
The 1$\sigma$ fluctuations in the depleted volume are approximately~10\,\%.

While the boundaries in $r$ are well defined, the lower boundaries in $z$ are less clearly resolved.
These result from falsely-validated events with large reconstruction errors, where the hit sequence in the camera was misreconstructed.
At \mbox{$r = 0\,\si{\milli\meter}$}, around half of the $z$-range of the detector is undepleted at \mbox{$V_B = -900\,\si{\volt}$}, and more than 75\,\% is undepleted for $V_B$ below \mbox{$-150\,\si{\volt}$}.
This results in a lack of events misreconstructed into the depleted volumes underneath the undepleted volumes. 
As event rates are generally higher at large $z$, see Fig.~\ref{fig:rate_depleted}, the contamination with such events is expected to be higher for bins at lower $z$. 
This explains why the relative efficiencies for volumes above the undepleted volumes do not show such a significant effect. 
Nevertheless, the shrinkage of the undepleted volume with $V_B$ is clearly visible also in~$z$.

\section{Determination of the impurity density profile from the Compton images}
\label{sec:detimp}

An empirically motivated~\mbox{\cite[p.\,105]{Hagemann2024}} parameterization for the radial dependence of the impurity density profile in a germanium detector was found to be a hyperbolic-tangent parameterization~\mbox{\cite{Abt2023,Darken2006}}. 
Assuming a linear $z$-dependence, this results in the following parameterization for the impurity density profile, $\rho$,~of
\begin{linenomath}
\begin{align} 
    \rho(r,z) = \bigg( &\underbrace{\rho_\infty + \Big(\rho_\text{in} \cdot (1  - \alpha \cdot \frac{r}{R}) - \rho_\infty \Big)}_{\textstyle \text{Linear $r$-dependence}}\,\cdot \notag{}\\
    &\underbrace{\frac{1}{2} \Big( \tanh(- \frac{r - r_0}{\lambda}) + 1 \Big)}_{\textstyle \text{tanh: center  at $r = r_0$}} \bigg)
    \cdot 
    \underbrace{ \Big( 1 + \beta \cdot \frac{z}{H} \Big)}_{\textstyle\makebox(0,0)[c]{\qquad Linear $z$-dependence}}~. \label{eq:tanh}
\end{align}
\end{linenomath}

Here, \mbox{$R = 37.25\,\si{\milli\meter}$} is the radius and \mbox{$H=39.5\,\si{\milli\meter}$} is the height of the detector.
The parameters $\rho_\infty, \rho_\text{in}, \alpha, r_0, \lambda$ and $\beta$ represent the impurity density at \mbox{$r \rightarrow \infty$} and \mbox{$r=0\,\si{\milli\meter}$}, a linear $r$ dependence, the position of the hyperbolic-tangent component and its strength, and the relative increase in impurity density from bottom to top, respectively.
While the absolute values provided by the manufacturer may be subject to large systematic uncertainties, the 12.1\,\% increase from bottom to top is expected to provide a reasonable estimate for $\beta$.
For this small value of $\beta$, simulations showed that the exact choice of the $z$-parameterization has a negligible influence on the other simulated observables, as long as it is monotonic.
Thus, $\beta$ was fixed to 12.1\,\% to ensure a robust fit to extract the $r$-dependence of the impurity density profile.

A fit across all bias voltages was performed by using predictions from the simulation and comparing them to data.
The output of the simulation was mapped onto the measured voxels. 
The scaling factors $f$ introduced in section~\ref{sec:sim} for each voxel represent the relative efficiencies for a perfect measurement. 
However, as the measurement was not perfect, a threshold was set and the voxel was counted as depleted, i.e.\;\mbox{$d=1$}, or undepleted, i.e.\;\mbox{$d=0$}, if the measured efficiency was above or below a certain percentage.
The default threshold was 50\,\%.
The loss function
\begin{linenomath}
\begin{equation} \label{eq:loss}
    L = \sum_{V_B} \sum_{\substack{\text{voxels with}\\ z > 10\,\si{\milli\meter}}}(f(V_B)-d(V_B))^2 
\end{equation}
\end{linenomath}
was minimized with parameters restricted such that the condition \mbox{$V_D = -1275\,\si{\volt}$} was fulfilled.
The spatial restriction to voxels with \mbox{$z > 10\,\si{\milli\meter}$} in the loss function mitigates systematic biases resulting from the reconstruction artifacts explained in section~\ref{sec:compton_images}, which effectively reduce count rates underneath undepleted volumes and cause some of them to be incorrectly classified as undepleted.

\begin{figure*}[!tb]
    \centering
    \bigskip
    \includegraphics{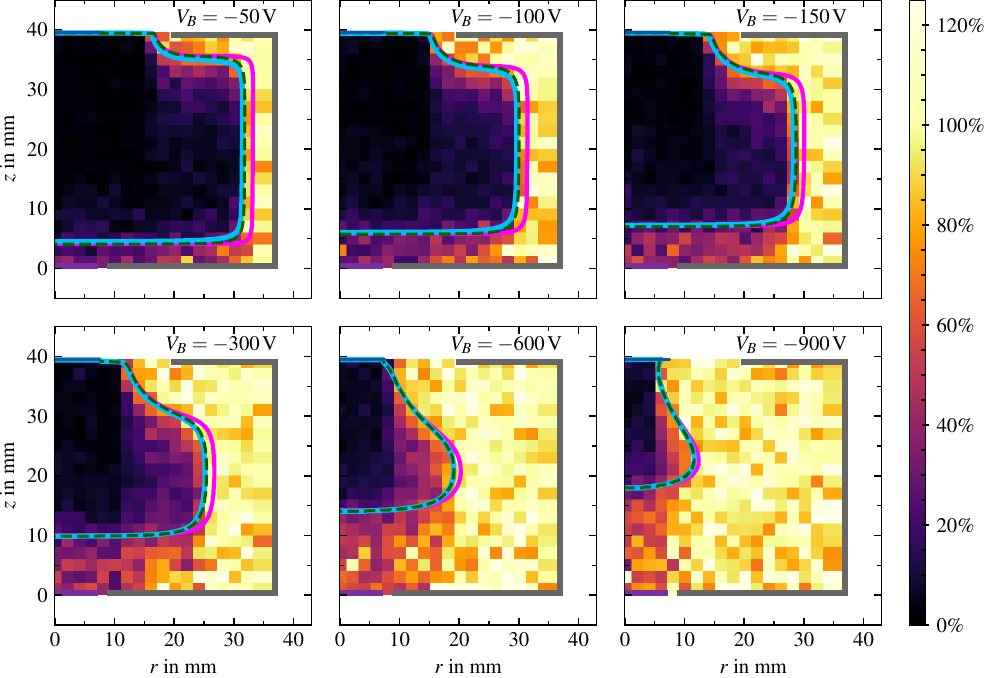} 
    \caption{
    Measured relative efficiencies from Fig.~\ref{fig:bubbles_undepleted}, overlaid with predictions using the impurity profile from the best fit (blue line) according to Eq.~\eqref{eq:tanh} for a threshold of 50\,\% to identify a voxel as depleted.
    Also shown are the predictions using the impurity values provided by the manufacturer scaled to 89\,\% (magenta line) without a radial dependence of the impurity density profile and the predictions from the result of a fit to the capacitance curve (dashed line), see section~\ref{sec:cap}.
    }
    \label{fig:bubbles_fits}
\end{figure*}

\begin{figure}[!tbhp]
\medskip
    \begin{overpic}{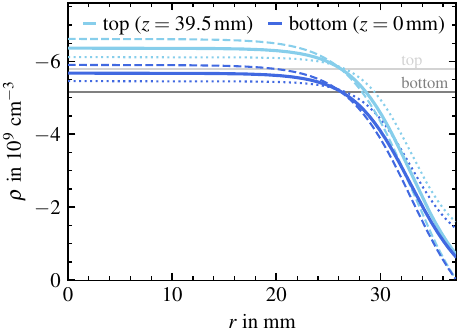} 
    \put(18,51){\makebox(0,0)[tl]{
        \begin{tabularx}{0.47\linewidth}{@{}>{\raggedright\arraybackslash}Xr@{\,}l@{}} \hline
            Parameter & Value \\ \hline
            $\rho_\text{in}$  & $(-5.68\pm0.22)\cdot10^{9}$ & cm$^{-3}$ \\
            $\rho_\infty $ & $(0.42${\raisebox{0.4ex}{\tiny$^{+1.06}_{-1.10}$}}$)\cdot10^{9}$ & cm$^{-3}$ \\
            $\alpha$   & $1.20${\raisebox{0.4ex}{\tiny$^{+0.16}_{-0.13}$}} & \% \\
            $r_0$      & $32.9${\raisebox{0.4ex}{\tiny$^{+0.3}_{-0.4}$}} & mm \\ 
            $\lambda$  & $5.56${\raisebox{0.4ex}{\tiny$^{+0.62}_{-0.92}$}} & mm \\
            $\beta$~(fixed) & 12.1 & \% \\
            \hline
        \end{tabularx}
    }}
    \end{overpic}
    \vspace{-.25\baselineskip}
    \caption{
    Impurity density profile of the p-type segBEGe detector obtained from fits to the Compton images, using the parameterization in Eq.\;\eqref{eq:tanh}. 
    The lines depict the best-fit results for threshold values of 50\,\% (solid), 40\,\% (dotted) and 60\,\% (dashed).
    The parameters are listed for the default threshold of 50\,\%.
    The variations of 10\,\% in the threshold values account for the 1$\sigma$-fluctuations of the relative efficiencies in the depleted regions, see Fig.~\ref{fig:bubbles_undepleted}.
    The light and dark gray horizontal lines depict the scaled manufacturer values at the top and bottom, respectively.
    }
    \label{fig:bestfit-bubble}
    \vspace{-\baselineskip}
\end{figure}

The predictions for the borders of the undepleted volumes for the different voltages are shown in Fig.~\ref{fig:bubbles_fits}. 
The predictions for the scaled manufacturer values are also shown as well as predictions from a capacitance measurement, see section~\ref{sec:cap}.

The impurity density profile as determined from the fit to the undepleted volumes is depicted in Fig.~\ref{fig:bestfit-bubble}.
The profile is provided for the top and the bottom of the detector. 
It is shown for the default threshold at 50\,\%, as well as for a threshold at 40\,\% and at 60\,\%, to identify depleted voxels. 
This variation of 10\,\% in the value of the threshold was chosen to define an envelope that accounts for the 10\,\% statistical fluctuations observed in the relative efficiency within the depleted volumes, see Fig.~\ref{fig:bubbles_fits}.
The lines intersect at \mbox{$r\approx26\,\si{\milli\meter}$}. 
This is a result of the restriction that the measured full-depletion voltage has to be reproduced.
Also provided are the parameters as defined in Eq.\;\eqref{eq:tanh} for the default threshold of~50\,\%.
The uncertainties on these parameters reflect the full range spanned when varying the threshold between 40\,\% and 60\,\% to account for $1\sigma$ statistical fluctuations in the relative efficiency.

The impurity density is almost constant for \mbox{$r\lesssim22\,\si{\milli\meter}$} while it drops rapidly for \mbox{$r\gtrsim22\,\si{\milli\meter}$}.
The probed volume extends to \mbox{$r\approx30\,\si{\milli\meter}$} corresponding to \mbox{$V_B=-50\,\si{\volt}$}, see Fig.~\ref{fig:bubbles_fits}. 
Anything beyond is an extrapolation of the fit. 
The probed volume ends \mbox{$\approx7\,\si{\milli\meter}$} away from the edge of the active volume of the detector, where the n-type lithium contact layer, see section~\ref{sec:det}, enters the calculations as \mbox{$0.75\,\si{\milli\meter}$} thick.
This layer does not have a sharp edge but certainly has no influence to radii as low as 
\mbox{$22\,\si{\milli\meter}$}, where the impurity density starts to decrease. It drops by about one order of magnitude
from \mbox{$r\approx22\,\si{\milli\meter}$} to the beginning of the lithium contact.

\begin{figure}[!b]
    \centering
    \begin{overpic}[scale = 0.85, trim={0 7mm 0 1mm}, clip]{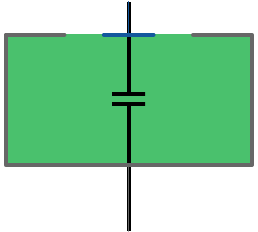} 
    \normalsize
    \put(60,35){\makebox(0,0)[l]{$C_d$}}
    \put(100,69){\makebox(0,0)[r]{$\vert V_B \vert > \vert V_D \vert$}}
    \put(50,-10){\makebox(0,0){\bfseries\large (a)}}
    \end{overpic}
    \hfill
    \begin{overpic}[scale = 0.85, trim={0 7mm 0 1mm}, clip]{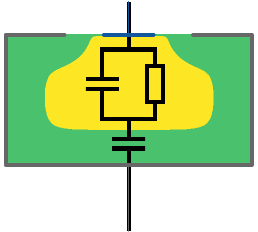} 
    \normalsize
    \put(100,69){\makebox(0,0)[r]{$\vert V_B \vert < \vert V_D \vert$}}
    \put(60,18){\makebox(0,0)[l]{$C_d$}}
    \put(66,40.5){\makebox(0,0)[l]{$R_u$}}
    \put(31.5,40.5){\makebox(0,0)[r]{$C_u$}}
    \put(50,-10){\makebox(0,0){\bfseries\large (b)}}
    \end{overpic}
    \bigskip
    \caption{
    Model used for the simulation of the capacitance of the p-type segBEGe detector for a (a)~fully-depleted and (b)~partially-depleted detector. 
    The depleted volume is shown in green, the undepleted volume is shown in yellow.
    }
    \label{fig:equivalent-CV-circuit}
\end{figure}

The fit parameters exhibit low overall correlation.
The absolute density scales $\rho_\text{in}$ and $\rho_\infty$ are decoupled from each other and only weakly correlated with the geometric parameters $\alpha$, $r_0$ and $\lambda$.
The strongest correlations are seen between $\alpha$ and $r_0$, as well as between $\lambda$ and $r_0$. 
These geometric correlations arise because the analysis is restricted to the physical volume of the detector, i.e.\;\mbox{$0 < r < R$}.
Within this region, shifting $r_0$ towards larger values can be partially compensated by a simultaneous increase in both $\alpha$ and $\lambda$.
Minor correlations are present between the geometric parameters and $\rho_\text{in}$ to reproduce the measured depletion voltage of \mbox{$V_D = -1275\,\text{V}$}, but they do not affect the fit stability.

Overall, the simulation with the impurity density profile as determined with a 50\,\% threshold fits the outer edge in $r$ of the undepleted volumes very well. 
The predictions from the manufacturer's values scaled to 89\,\% with no radial dependence are also shown in Fig.~\ref{fig:bubbles_fits}.
These predictions overestimate the radial extent of the undepleted volume for all \mbox{$\vert V_B\vert < \vert V_D\vert$}.
This cannot be avoided without introducing a radial dependence of the impurity density.
A strong decrease of the impurity density towards the edge of the detector is needed to obtain a good description of the undepleted volumes for all $V_B$ simultaneously.

\section{Comparison to capacitance measurements} 
\label{sec:cap}

The geometry of the depleted volume defines the capacitance of the detector, $C_d$.
Thus, the impurity density profile can also be inferred from capacitance measurements.

Figure~\ref{fig:equivalent-CV-circuit} demonstrates how the detector has to be modeled when it is fully or partially depleted. 
The depleted volume is responsible for~$C_d$. 
The undepleted volume is represented by a capacitance~$C_u$ and a resistance~$R_u$ connected in parallel. 
Given an impurity density profile, the values of $C_d$ for a set of $V_B$ can be calculated with \SSD{} using the gradients of the weighting potentials~\cite{Smolic2021}.

\begin{figure}[!b]
    \centering
    \medskip
    \definecolor{segBEGeblue}{RGB}{19,88,157}
\definecolor{segBEGegray}{RGB}{103,103,103}
\begin{tikzpicture}[scale = 0.29, line width = 1]
    \fill [fill = lightgray] (-37.25mm,0) -- (-37.25mm,-39.5mm) arc (180:360:37.25mm and 1cm) -- (37.25mm,0) arc (0:180:37.25mm and 1cm);
    \filldraw [line width=0.2, fill = segBEGegray] (28:19.5mm and 0.523cm) -- (28:37.25mm and 10mm) arc (28:0:37.25mm and 10mm) -- ([shift={(0:37.25mm and 10mm)}]0,-39.5mm) arc (0:-28:37.25mm and 10mm) -- (-28:37.25mm and 10mm) -- (-28:19.5mm and 0.523cm) arc (-28:28:19.5mm and 0.523cm) -- cycle;
    \filldraw [line width=0.2, fill = segBEGegray] (-31:19.5mm and 0.523cm) -- (-31:37.25mm and 10mm) -- (-31:37.25mm and 10mm) -- ([shift={(-31:37.25mm and 10mm)}]0,-39.5mm) arc (-31:-89:37.25mm and 10mm) -- (-89:37.25mm and 10mm) -- (-89:19.5mm and 0.523cm) arc (-89:-31:19.5mm and 0.523cm) -- cycle;
    \filldraw [line width=0.2, fill = segBEGegray] (-91:19.5mm and 0.523cm) -- (-91:37.25mm and 10mm) -- (-91:37.25mm and 10mm) -- ([shift={(-91:37.25mm and 10mm)}]0,-39.5mm) arc (-91:-149:37.25mm and 10mm) -- (-149:37.25mm and 10mm) -- (-149:19.5mm and 0.523cm) arc (-149:-91:19.5mm and 0.523cm) -- cycle;
    \filldraw [line width=0.2, fill = segBEGegray] (152:19.5mm and 0.523cm) -- (152:37.25mm and 10mm) arc (152:180:37.25mm and 10mm) -- ([shift={(180:37.25mm and 10mm)}]0,-39.5mm) arc (180:208:37.25mm and 10mm) -- (208:37.25mm and 10mm) -- (208:19.5mm and 0.523cm) arc (208:152:19.5mm and 0.523cm) -- cycle;
    \filldraw [line width=0.2, fill = segBEGegray] (149:19.5mm and 0.523cm) -- (149:37.25mm and 10mm) arc (149:91:37.25mm and 10mm) -- (91:19.5mm and 0.523cm) arc (91:149:19.5mm and 0.523cm) -- cycle;
    \filldraw [line width=0.2, fill = segBEGegray] (31:19.5mm and 0.523cm) -- (31:37.25mm and 10mm) arc (31:89:37.25mm and 10mm) -- (89:19.5mm and 0.523cm) arc (89:31:19.5mm and 0.523cm) -- cycle;
    \draw[line width=0.2] (0,0) ellipse (37.25mm and 1cm);
    \draw[line width=0.2, fill = segBEGeblue] (0,0) ellipse (7.5mm and 0.2 cm);

    \draw [line width=2] (-50mm,23mm) -- (-50mm,40mm) -- (50mm,40mm) -- (50mm,23mm);
    \draw [line width=2] (-50mm,17mm) -- (-50mm,-70mm) --(-3mm, -70mm);
    \draw [line width=2] (3mm,-70mm) -- (50mm,-70mm) -- (50mm,17mm);
    \draw  (62mm,-105mm) -- (78mm,-105mm);
    \draw  (65mm,-108mm) -- (75mm,-108mm);
    \draw  (68mm,-111mm) -- (72mm,-111mm);
    \node at (-50mm,46mm)  {\makebox(0,0)[l]{K2 cryostat}};

    \draw [line width=2] (37.25mm,-19.75mm) -- (43mm,-19.75mm) -- (43mm, -55mm) -- (0,-55mm);
    \draw [line width=2] ([shift={(300:37.25mm and 1cm)}]0,-19.75mm) -- (18.625mm,-55mm) -- (0, -55mm);
    \draw [line width=2] ([shift={(240:37.25mm and 1cm)}]0,-19.75mm) -- (-18.625mm,-55mm) -- (0,-55mm) -- (0,-90mm);
    \draw [line width=2] (-37.25mm,-19.75mm) -- (-43mm,-19.75mm) -- (-43mm, -55mm) -- (0,-55mm);
    \draw [line width=2] (-120mm,-90mm) -- (70mm,-90mm);
    \draw [line width=2] (50mm,-60mm) -- (70mm,-60mm) -- (70mm,-105mm);
    \draw [color = segBEGegray] (37.25mm,-19.75mm) -- (43mm,-19.75mm) -- (43mm, -55mm) -- (0,-55mm);
    \draw [color = segBEGegray] ([shift={(240:37.25mm and 1cm)}]0,-19.75mm) -- (-18.625mm,-55mm) -- (0,-55mm) -- (0,-90mm);
    \draw [color = segBEGegray] ([shift={(300:37.25mm and 1cm)}]0,-19.75mm) -- (18.625mm,-55mm) -- (0, -55mm);
    \draw [color = segBEGegray] (-37.25mm,-19.75mm) -- (-43mm,-19.75mm) -- (-43mm, -55mm) -- (0,-55mm);
    \draw [color = segBEGegray] (-120mm,-90mm) -- (70mm,-90mm);
    \draw [color = segBEGegray] (50mm,-60mm) -- (70mm,-60mm) -- (70mm,-105mm);
    
    \draw  [line width=2] (0,0) -- (0,20mm);
    \draw  [line width=2] (-120mm,20mm) -- (67.5mm,20mm);
    \draw  [line width=2] (72.5mm,20mm) -- (130mm,20mm);
    \draw  [color = segBEGeblue] (0,0) -- (0,20mm);
    \draw  [color = segBEGeblue] (-120mm,20mm) -- (67.5mm,20mm);
    \draw  [color = segBEGeblue] (72.5mm,20mm) -- (130mm,20mm);

    \draw  [fill = white] (-110mm,13mm) -- (-75mm,13mm) -- (-75mm,27mm) -- (-110mm,27mm) -- cycle;
    \draw  (-132mm,20mm) circle (12mm);
    \draw  (-140mm,20mm) -- (-124mm,20mm);
    \draw  (-65mm, 23mm) -- (-65mm,40mm) -- (-150mm,40mm) -- (-150mm,0mm) -- (-65mm,0mm) -- (-65mm,17mm);
    \node  at (-117mm,32mm) {$V_B$};
    \node  at (-92.5mm,20mm) {$R_B$};
    \node at (-150mm,46mm)  {\makebox(0,0)[l]{High-voltage supply}};

    \draw  [fill=white] (-50mm,-83mm) -- (-15mm,-83mm) -- (-15mm,-97mm) -- (-50mm,-97mm) -- cycle;
    \node  at (-32.5mm,-90mm) {$R_1$};
    \draw  [fill=white] (50mm,-83mm) -- (15mm,-83mm) -- (15mm,-97mm) -- (50mm,-97mm) -- cycle;
    \node  at (32.5mm,-90mm) {$R_2$};
    \draw  [fill=white] (-110mm,-83mm) -- (-75mm,-83mm) -- (-75mm,-97mm) -- (-110mm,-97mm) -- cycle;
    \node  at (-92.5mm,-90mm) {$R_\text{PG}$};
    \node  at (-116mm,-77mm) {$V_\text{PG}$};
    \node  at (0mm,-98mm) {$V_\text{in}$};
    \draw  (-132mm,-90mm) circle (12mm);
    \draw  (-140mm,-90mm) sin (-136mm,-87mm) cos (-132mm,-90mm) sin (-128mm,-93mm) cos (-124mm,-90mm);
    \draw  (-65mm, -93mm) -- (-65mm,-110mm) -- (-150mm,-110mm) -- (-150mm,-70mm) -- (-65mm,-70mm) -- (-65mm,-87mm);
    \node at (-150mm,-64mm)  {\makebox(0,0)[l]{Pulse generator}};

    \draw  (67.5mm,10mm) -- (67.5mm,30mm);
    \draw  (72.5mm,10mm) -- (72.5mm,30mm);
    \node  at (70mm,1mm) {$C_C$};

    \draw  [fill=white] (92mm,10mm) -- (92mm,30mm) -- (110mm,20mm) -- cycle;
    \node at (101mm, 5mm) {\makebox(0,0)[t]{\shortstack[c]{Charge-\\ sensitive \\ preamplifier}}};

    \draw [fill = segBEGeblue] (130mm,20mm) circle (1.5mm);
    \draw  (127.172mm,17.172mm) arc (-135:135:4mm);
    \node at (130mm,31mm) {\makebox(0,0)[c]{ADC}};
\end{tikzpicture}
    \caption{
    Schematic of the pulser system used to measure capacitances of the p-type segBEGe detector in the K2 cryostat.
    A pulse generator is used to inject voltage pulses of known amplitude $V_\text{in}$ to the segments.
    The increase in charge on the core contact is read out using a charge-sensitive preamplifier readout, with AC coupling via a coupling capacitor $C_C = 1.2\,\si{\nano\farad}$ to block the bias voltage $V_B$.
    }
    \label{fig:electric_circuit}
\end{figure} 

The capacitance-voltage (CV) curve of a detector can be measured using a pulser system~\mbox{\cite{Birkenbach2011,Abt2023}}.
Figure~\ref{fig:electric_circuit} shows the pulser system that was directly installed on the K2 cryostat housing the p-type segBEGe detector.
The system was used to apply voltage step pulses with known amplitude $V_\text{in}$ to the shorted segments and to measure the increase of charge $Q_\text{out}$ on the core contact.
The detector capacitance $C_d$ is connected in series with the coupling capacitor $C_C$, resulting in a total capacitance $C_\text{tot}$ of 
\begin{linenomath}
\begin{equation}
    \dfrac{1}{C_\text{tot}} = \dfrac{1}{C_d} + \dfrac{1}{C_C} \quad \Rightarrow \quad C_\text{tot} = \dfrac{C_d \cdot C_C}{C_d + C_C}~. \label{eq:Ctot}
\end{equation}
\end{linenomath}
The values of $C_d$ were determined by extracting $Q_\text{out}$ from the charge pulses and correcting the resulting values for \mbox{$C_\text{tot} = -Q_\text{out} / V_\text{in}$} for $C_C = 1.2\,\si{\nano\farad}$ using Eq.\;\eqref{eq:Ctot}.
For a fully-depleted detector, the charge pulses from pulser events quickly reach their final amplitude of $Q_\text{out} = -V_\text{in} C_\text{tot}$.
For a partially-depleted detector, the RC response of the undepleted volume leads to a slower exponential approach of the charge pulse to its final amplitude~\cite{Haller1972}:
\begin{linenomath}
\begin{equation} \label{eq:Haller1972}
    Q(t) = 
    \underbrace{
    -V_\text{in}\,C_\text{tot}}_{\textstyle = Q_\text{out}} \!\left( 1 - \dfrac{C_\text{tot}}{C_u + C_\text{tot}}\exp\!\left(\!-\dfrac{t}{R_u (C_u + C_\text{tot})}\!\right)\!\right)~,
\end{equation}
\end{linenomath}
where the time constant $\tau_u = R_u (C_u + C_\text{tot})$ is determined by the geometry of the undepleted volume.

The capacitance was measured at 98\,voltages down to \mbox{$V_B=-120\,\si{\volt}$}, which took about 8\,hours.
Measurements below \mbox{$-120\,\si{\volt}$} were outside of the operating range of the ADC. 
Thus, the outer region of the detector could not be probed to quite as large $r$ as with the Compton images. 

\begin{figure}[!tb]
    \centering
    \includegraphics{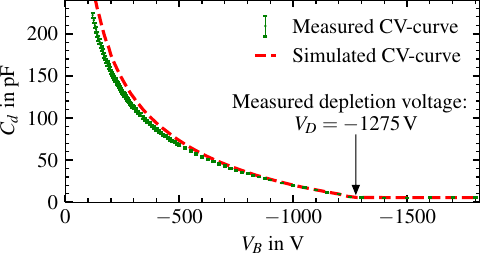} 
    \caption{
    Measured capacitance values (points with error bars) together with the predicted CV curve using the impurity density values from the manufacturer scaled to 89\,\% to match $V_D$  (dashed red line).
    }
    \label{fig:measured=CV}
\end{figure}

The measured CV curve is shown in Fig.~\ref{fig:measured=CV}.
The uncertainties on $C_d$ are dominated by a $1.7\%$ systematic uncertainty in the input-voltage amplitude $V_\text{in}$, which is identical across all $V_B$ and therefore highly correlated between capacitance values~\mbox{\cite[p.\,103]{Hagemann2024}}.
The statistical uncertainty of 0.17\% was estimated separately from the fluctuations in the full-depletion capacitance above $V_D$.
Also shown in Fig.~\ref{fig:measured=CV} is the prediction based on the manufacturer values scaled to 89\,\% to match $V_D$.
The detector capacitance $C_d$ has a constant value for $\vert V_B\vert \ge \vert V_D \vert$.
While $V_D$ is correctly predicted for the scaled values from the manufacturer, the predictions for $C_d$ below $V_D$ are all above the measured values with the deviation growing with reduced $V_B$.
This can, again, not be explained unless a radial dependence of the impurity density profile is introduced.

\begin{figure}[!tb]
    \centering
    \includegraphics{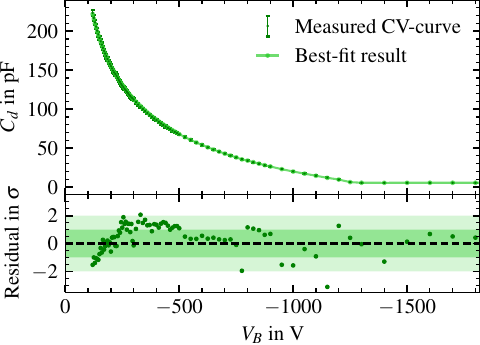} 
    \caption{
    Comparison of the simulated CV curve using the best-fit result according to Eq.~\eqref{eq:tanh} to the measured capacitance values. 
    Also shown are the residuals for the data (green points) compared to the $1\sigma$ and $2\sigma$ uncertainty bands of the predictions.
    }
    \label{fig:CV}
\end{figure}

A maximum likelihood fit based on \textit{Optim.jl}~\cite{Optim.jl} was performed to obtain the parameters of Eq.~\eqref{eq:tanh} with a fixed \mbox{$\beta = 12.1\,\%$} as used for the fit to the Compton images plus the requirements of \mbox{$\alpha < 1$}, \mbox{$0 < r_0 < R$} and \mbox{$\lambda > 0$}. 
The restriction on $V_D$ could be omitted as $V_D$ is determined from the voltage at which the curve becomes a constant and was extracted correctly automatically.
Figure~\ref{fig:CV} shows the data together with the predicted CV curve using the best-fit impurity density profile.  
Also shown are the residuals of the data with respect to the $1\sigma$ and $2\sigma$ uncertainty bands on the prediction.
The predictions are slightly too low between \mbox{$V_B\approx-200\,\si{\volt}$} and \mbox{$V_B\approx-500\,\si{\volt}$} and become too high at lower voltages.
These deviations point to minor limitations of the factorizable parameterization in Eq.\;\eqref{eq:tanh}, which cannot model correlated $r$- and $z$-dependencies or localized impurity variations.
Interestingly, the radial extent of the undepleted region at around \mbox{$V_B = -300\,\si{\volt}$} of $r \approx 25\,\si{\milli\meter}$ shown in Fig.~\ref{fig:bubbles_fits} coincides with the radius at which $\rho$ was seen to decrease.

\begin{figure}[!tb]
    \medskip
    \begin{overpic}{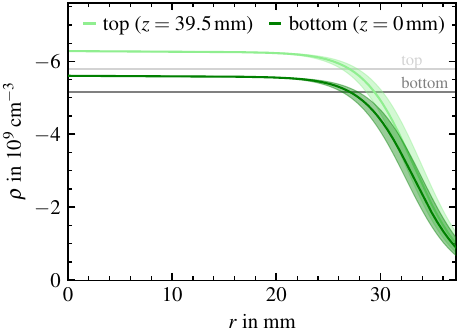} 
    \put(18,51){\makebox(0,0)[tl]{
        \begin{tabularx}{0.47\linewidth}{@{}>{\raggedright\arraybackslash}Xr@{\,}l@{}} \hline
            Parameter & Value \\ \hline
            $\rho_\text{in}$  & $(-5.61\pm0.01)\cdot10^{9}$ & cm$^{-3}$ \\
            $\rho_\infty$ & $(-0.006\pm0.001)\cdot10^{9}$ & cm$^{-3}$ \\
            $\alpha$   & $0.81\pm0.01$ & \% \\
            $r_0$      & $33.2\pm0.4$ & mm \\
            $\lambda$  & $4.86\pm0.02$ & mm \\
            $\beta$~(fixed) & 12.1 & \% \\
            \hline
        \end{tabularx}
    }}
    \end{overpic}
    \caption{
    Impurity density profile of the p-type segBEGe detector obtained from a maximum likelihood fit to the measured CV-curve using the parameterization in Eq.~\eqref{eq:tanh}. 
    The bands indicate $2\sigma$-uncertainties on the data. 
    Indicated as dashed lines are the linear functions describing the impurity density profile in the central part of the detector.
    The light and dark gray horizontal lines depict the scaled manufacturer values at the top and bottom, respectively.
    }
    \label{fig:bestfit-CV}
\end{figure}

\begin{figure*}[!b]
    \medskip
    \centering
    \begin{overpic}{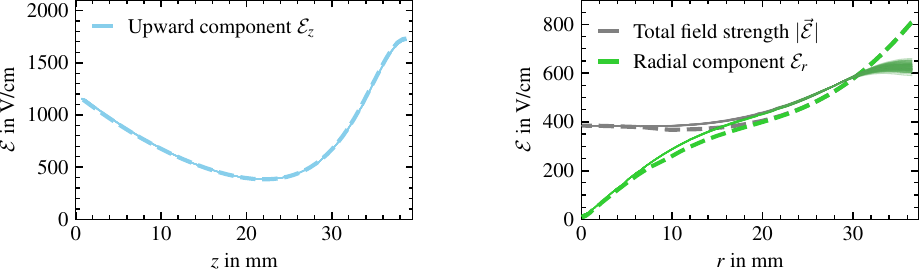}
    \bfseries\large
    \put(26.5,-2){\makebox(0,0){(a)}}
    \put(82,-2){\makebox(0,0){(b)}}
    \end{overpic}
    \bigskip
    \addtocounter{figure}{1}
    \caption{
    (a) Upward component of the electric field, $\mathcal{E}_z(z)$, in the center of the detector at \mbox{$r = 0\,\si{\milli\meter}$}, and (b) radial component, $\mathcal{E}_r(r)$, and total electric field strength $\vert\vec{\mathcal{E}}(r)\vert$ at \mbox{$z = 21.5\,\si{\milli\meter}$}, for selected regions highlighted in Fig.~\ref{fig:e_field_2D}.
    Predictions from the best-fit result from Fig.~\ref{fig:bestfit-CV} are shown as solid lines.
    Predictions based on the manufacturer values, scaled to 89\,\% and assuming no radial dependence, are shown as dashed lines.
    }
    \addtocounter{figure}{-2}
    \label{fig:e_field_lines}
\end{figure*}

The net impurity density profile as determined by the best fit is shown in Fig.~\ref{fig:bestfit-CV}. 
It is very similar to the profile as determined from the fit to the undepleted volumes shown in Fig.~\ref{fig:bestfit-bubble}. 
The parameter uncertainties were calculated using the Laplace approximation and are purely statistical.
In contrast, the envelope in Fig.~\ref{fig:bestfit-bubble} is determined by varying the depletion threshold and also account for spatial reconstruction systematics.
The values for $\rho_\text{in}$ and $\rho_\infty$ agree within uncertainties. 
The fit uncertainties are smaller for the capacitance fit for these densities. 
It should, however, be remembered that $\rho_\infty$ comes from an extrapolation beyond the range of the data and that the capacitance fit does not probe to as large $r$ as the fit to the Compton images.
Both fits indicate that the impurity density remains almost constant for \mbox{$r\lesssim22\,\si{\milli\meter}$}. 
The values for $r_0$ agree within uncertainties, where both fits show similar uncertainties. 
This means that the downturn of impurity density sets in around \mbox{$r \approx 22\,\si{\milli\meter}$} for both fits. 
The values of $\lambda$ are also compatible with large uncertainties from the fit to the Compton images.
While the values for $\alpha$ exhibit a formal $3\sigma$ discrepancy due to tiny statistical uncertainties of $\pm 0.01\%$ from the capacitance fit, both fits point to the same physical picture with a small linear $r$-gradient near 1\%.
Overall, despite their fundamentally different sources of systematic uncertainties, the capacitance fit and the fit to the Compton images show excellent agreement.

\section{Implications for the electric field}

The refined knowledge on the radial impurity density profile directly impacts the calculation of the electric field in the fully-depleted detector.
Figure~\ref{fig:e_field_2D} depicts the electric field, $\vec{\mathcal{E}}(\vec{r})$, calculated at a bias voltage of \mbox{$V_B = -3000\,\si\volt$} using the best-fit result from the capacitance measurements.
To evaluate the results, this electric field is compared against the baseline shown in Fig.~\ref{fig:epot-components}, obtained from the manufacturer values scaled to 89\,\% and assuming no $r$-dependence.

\newpage

\begin{figure}[!tbh]
    \centering
    \null\quad\includegraphics{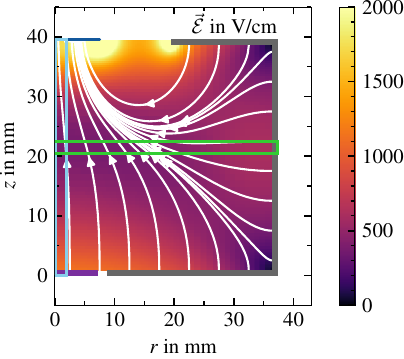}
    \caption{
    Electric field in the p-type segBEGe calculated using the impurity density as determined from the best-fit result from the capacitance measurements shown in Fig.~\ref{fig:bestfit-CV} and a bias voltage of \mbox{$V_B = -3000\,\si{\volt}$}.
    Electric field profiles in the highlighted boxes are shown in Fig.~\ref{fig:e_field_lines}.
    }
    \label{fig:e_field_2D}
    \addtocounter{figure}{1}
\end{figure}

The radial and upward field components $\mathcal{E}_r$ and $\mathcal{E}_z$ were extracted along the representative horizontal and vertical lines highlighted in Fig.~\ref{fig:e_field_2D} and are shown in Fig.~\ref{fig:e_field_lines}.
The uncertainties on the impurity density profile were propagated by sampling parameter sets \mbox{($\rho_\text{in}$, $\rho_\infty$, $\alpha$, $r_0$, $\lambda$)} from the Hessian covariance matrix.
The resulting field variations are indicated by the uncertainty bands in Fig.~\ref{fig:e_field_lines}.

Along the central axis of the detector at \mbox{$r = 0\,\si{\milli\meter}$}, the electric field points upwards, i.e.\;$\mathcal{E}_z = \vert\vec{\mathcal{E}}\vert$, see Fig.~\ref{fig:e_field_lines}a.
It features a minimum value of around $380\,\si{\volt/\centi\meter}$ at \mbox{$z = 21.5\,\si{\milli\meter}$}, reaches its maximum of around $1725\,\si{\volt/\centi\meter}$ near the core contact and increases towards the bottom surface to around $1150\,\si{\volt/\centi\meter}$ at \mbox{$z = 0\,\si{\milli\meter}$}.
This vertical profile shows no noticeable difference from the baseline electric~field.

In contrast, significant deviations from the baseline electric field appear in the radial component $\mathcal{E}_r$ along the horizontal profile at \mbox{$z = 21.5\,\si{\milli\meter}$}, see Fig.~\ref{fig:e_field_lines}b. 
At this $z$, the electric field points inwards for \mbox{$r > 20\,\si{\milli\meter}$}, i.e.\;\mbox{$\mathcal{E}_r \approx \vert\vec{\mathcal{E}}\vert$}.
The updated electric field is almost constant at \mbox{$400\,\si{\volt/\centi\meter}$} for \mbox{$r \lesssim 18\,\si{\milli\meter}$} and increases to only \mbox{$(625\pm25)\,\si{\volt/\centi\meter}$} near the outer surface. 
This differs substantially from the baseline electric field which rises to above \mbox{$750\,\si{\volt/\centi\meter}$} at the outer contact.

The reduced electric field affects the drift dynamics of charge carriers.
For events close to the outer surface, the initial inward drift of holes is slower than predicted assuming no $r$-dependence.
Likewise, the final outward drift of electrons towards the segments is also slowed down.
Because the modifications are spatially non-uniform, their net impact on pulse shape observables depends strongly on the spatial distribution of the interaction positions.
Thus, a simple generalized statement on the overall change in high-level pulse shape characteristics would be physically unjustified.

Aside from the electric field, pulse shape simulations also rely on accurate descriptions of charge drift mobilities and their anisotropies.
The precise electric field presented here marks the first step to decouple electric-field effects from these charge drift properties.
As the Compton Scanner provides each pulse with its reconstructed position, different mobility models can be tested by comparing simulated pulses directly to measured data.
The feasibility of using Compton Scanner data to extract charge carrier mobilities has already been demonstrated~\cite{Hagemann2024}, establishing a powerful framework to systematically improve the precision of pulse shape simulations for next-generation rare-event searches.

\section{Summary and discussion}

A novel method to directly image the undepleted volumes of a germanium detector at different bias voltages based on three-dimensional Compton imaging was introduced, and such Compton images were shown for a p-type Broad Energy germanium detector.
The predictions of undepleted volumes calculated with the open-source software package \SSD{} were fitted to the images.
The shape of the undepleted volumes could only be described if a radial dependence of the impurity density profile was assumed. 
The necessity to introduce such a radial dependence was confirmed by analyzing the capacitance-voltage curve of the detector. 
The impurity density profile in the center of the detector is almost constant up to a radius of approximately \mbox{$22\,\si{\milli\meter}$}, where it starts to decrease rapidly.
An extrapolation to the edge of the detector indicates that the net density of electrically active impurities might become very low.
This work demonstrates that three-dimensional Compton imaging of undepleted detector volumes is a powerful tool to obtain the impurity density profile of a germanium detector.

Rare-event searches use pulse shape analysis to identify background events.
Such an analysis can only be reliably developed from simulations for detectors with a known impurity density profile.
Capacitance-voltage measurements are relatively fast, taking about one day per detector, and should be performed down to the lowest achievable voltages to probe as close to the surface as possible.
However, measured capacitances represent only integrated values over the depleted volumes of the detector, limiting their ability to unambiguously determine the $r$- and $z$-components of the impurity density profile.

In contrast, three-dimensional Compton images of undepleted volumes can provide the complementary information required to disentangle the individual components of the impurity density profile.
This capability is critical because the radial dependence may vary between crystal growing facilities and detector types, and the assumption of a linear $z$-dependence might not hold for larger-height detectors.
Although scanning every detector in a large array is impractical due to scan times of several weeks, applying this technique to selected detectors is highly viable:
the feasibility of scanning multi-kilogram detectors has already been demonstrated, and the cost of a Compton Scanner as used to create the images presented in this paper is negligible compared to the overall cost of a large-scale germanium-based experiment.
For large detector arrays, the recommended strategy is to measure capacitance-voltage curves for all detectors, while taking three-dimensional Compton images of the undepleted volumes for selected detectors, representing each detector type and crystal growing facility.

\begin{acknowledgements}
We thank our former colleagues Dr.\ Lukas Hauertmann and Dr.\ Martin Schuster for helpful discussions and support.
\end{acknowledgements}

\bibliographystyle{spphys.bst}
\bibliography{main.bib}

\end{document}